\documentclass[showpacs,10pt,aps,prd,reprint,onecolumn,notitlepage,nofootinbib,superscriptaddress]{revtex4-1}

\usepackage{hyperref}
\usepackage{amsmath}
\usepackage{mathtools}
\usepackage{bm}
\usepackage{amssymb}
\usepackage{graphicx}
\usepackage[]{subfigure}
\usepackage{filecontents}
\usepackage[dvipsnames]{xcolor}

\hypersetup{
    bookmarks=true,         % show bookmarks bar?
    pdftoolbar=true,        % show Acrobats toolbar?
    pdfmenubar=true,        % show Acrobats menu?
    pdffitwindow=true,     %window fit to page when opened
    pdfstartview={FitH},     %fits the width of the page to the window
    pdftitle={My title},     %title
    pdfauthor={author},      %author
    pdfsubject={Subject},    %subject of the document
    pdfcreator={Creator},    %creator of the document
    pdfproducer={Producer},  %producer of the document
    pdfkeywords={keyword1} %{key2} {key3}, % list of keywords
    pdfnewwindow=true,       %links in new window
    colorlinks=true ,        %false: boxed links; true: colored links
    linkcolor=blue  ,        %color of internal links (change box color with linkbordercolor)
    citecolor=blue,      %color of links to bibliography
    filecolor=green,       %color of file links
    urlcolor=blue            %color of external links
}

\usepackage{yfonts}
\providecommand{\ed}{\mathrm{d}}

\providecommand{\ed}{\mathrm{d}}

\DeclareMathOperator{\arccot}{arccot}

\begin{document}

\title{{The role of elliptic integrals in calculating the gravitational lensing of a charged Weyl black hole surrounded by plasma}}
%\subtitle{Do you have a subtitle?\\ If so, write it here}

%\titlerunning{Short form of title}        % if too long for running head

\author{Mohsen Fathi}
\email{mohsen.fathi@postgrado.uv.cl}
\affiliation{Instituto de F\'isica y Astronom\'ia, Universidad de Valpara\'iso, Avenida Gran Breta\~na 1111, Valpara\'iso, Chile}

\author{J.R. Villanueva}
\email{jose.villanueva@uv.cl}
\affiliation{Instituto de F\'isica y Astronom\'ia, Universidad de Valpara\'iso, Avenida Gran Breta\~na 1111, Valpara\'iso, Chile}

%\authorrunning{Short form of author list} % if too long for running head

%\date{\today}
% The correct dates will be entered by the editor

\begin{abstract}
{In this paper, we mainly aim at highlighting the importance of (hyper-)elliptic integrals in the study of gravitational effects caused by strongly gravitating systems. For this, we study the application of elliptic integrals in calculating the light deflection as it passes a plasmic medium, surrounding a charged Weyl black hole. To proceed with this, we consider two specific algebraic ansatzes for the plasmic refractive index, and we characterize the photon sphere for each of the cases. This will be used further to calculate the angular diameter of the corresponding black hole shadow. We show that the complexity of the refractive index expressions, can result in substantially different types of dependencies of the light behavior on the spacetime parameters. }

\bigskip

\noindent{\textit{keywords}}: Elliptic integrals, light deflection, black holes

\end{abstract}
\maketitle

%%%%%%%%%%%%%%%%%%%%%%%%%%%%%
\section{Introduction}

{The equations of motion for particles travelling in the gravitational fields of massive objects, as formulated by general theory of relativity, have been receiving a rigorous attention ever since the advent of the theory. In fact, the approximate solutions to these equations, at the time, could pave the way in figuring out the trajectories of planets and light in the solar system and finally, lead to some observational evidences which confirmed general relativity's predictions (as asserted by Eddington in his famous book \cite{Eddington1920}).}

{However, the more delicate the experimental tests became, the more they raised the interest in obtaining exact solutions to the equations of motion. This necessitated employing advanced mathematical methods, mainly, because of the resultant differential equations appearing in the equations of motion, which tend to calculate the arc-lengths associated with the particle trajectories. Since the early attempts by Hagihara \cite{Hagihara:1931} and Darwin \cite{Darwin:1959,Darwin:1961} in obtaining and categorizing the particle orbits in the Schwarzschild spacetime, researchers have been employing different approaches to the computation of the arc-lengths swept by particle trajectories in gravitating systems. These approaches are, in general, based on manipulating elliptic integrals and the resultant elliptic functions, covering the Jacobi and the Weierstra\ss~elliptic functions, as the two most common forms. Ever since, the elliptic and hyper-elliptic functions have received a great deal of interest in analyzing the geodesic structure of massive and mass-less particles in black hole spacetimes \cite{Rauch:1994, Beckwith:2005, Hackmann:2008a, Hackmann:2008b, Hackmann:2008c, Hackmann:2008d, Kogan:2008a, Hackmann:2009a, Hackmann:2010a, Hackmann:2010b, Grunau:2011, Hackmann:2012a, Gibbons:2012, Hackmann:2014a, Gerardo:2014, Hackmann:2015a, DeFalco:2016, Barlow:2017, Jusufi:2018, Ghaffarnejad:2018, Villanueva:2018kem, Hsiao:2020} (also see Ref.~\cite{Vankov:2017}).}

{Along the same efforts, this research is dedicated to the application of elliptic integrals in calculating the gravitational lensing of light rays (mass-less particles) passing a particular black hole spacetime. The importance of this mathematical method becomes more highlighted, when the black hole is supposed to be merged in an inhomogeneous plasma, described by a coordinate-dependent refractive index. In fact, the usage of elliptic integrals in studying the light deflection in black hole spacetimes filled with plasma, has been dealt with for some regular black holes (see, for example, Ref.~\cite{Kogan:2017} for a good review). In this work, however, we try to get more insights to the abilities of the elliptic integrals in the calculation of the deflection angles of light ray trajectories, by choosing specific refractive ansatzes that are complicated enough, to be able to include a wide range of dependencies of the black hole surroundings on the horizon distances. This is done by considering two ansatzes for the plasmic refractive indices that are expressed as functions of the black hole horizons. The background spacetime under consideration, corresponds to a static spherically symmetric electrically charged black hole, proposed in Ref.~\cite{Payandeh:2012mj}, which is inferred from Weyl conformal theory of gravity. This black hole has also been recently examined in Refs.~\cite{Fathi:2020,Fathi:2020b,Fathi:2021a}, regarding the behavior of null and time-like geodesics passing its exterior geometry, where the elliptic functions played an important role in the determination of the mass-less and (charged) massive particle trajectories. The shadow structure of a rotating counterpart of this black hole has been also discussed in Ref.~\cite{Fathi:2021b}.}

{Beside calculating the deflection angle, in this paper, we also relate the aforementioned ansatzes to the black hole's photon sphere and shadow. The role of the elliptic integrals becomes more apparent in this regard, since without knowing the ability of the black hole in bending the light, one cannot talk about related features.}

{Accordingly, we organize this study as follows: In Sec.~\ref{ngse}, we first bring some fundamentals on optical gravity and establish a Hamiltonian formalism on the spacetime manifold's cotangent bundle. In this section, we also introduce the general formulation of the light's deflation angle. In Sec.~\ref{sec:nAnzats}, we propose two substantially different ansatzes as the plasmic refractive indices. Accordingly, we employ the method of elliptic integration to calculate the light deflection angle for each case and discuss their peculiarities. Furthermore, by calculating the radius of the photon spheres in either of the media with given refractions, in Sec.~\ref{sec:rph}, we stipulate to what extent the orbiting photons can approach the black hole without falling into its event horizon. In Sec.~\ref{sec:Nr} we put a small gap in our discussion to talk more about physical implications of the refractive plasma. To do this, we discuss the particle concentration in the region of casual connection outside the black hole event horizon, which is followed by making a comparison between the refractive media and a dark matter halo in the context of Navarro-Frenk-White density profile \cite{Navarro1995c,Navarro1996}. The significance of photon spheres is used in Sec.~\ref{sec:shadowRadius} to obtain the angular diameter of the black hole shadow in each of the mentioned plasmic media. The results also show some confinements on the impact parameter which is associated with the trajectories. Final notes are given in Sec.~\ref{sec:conclusions}.  }

%%%%%%%%%%%%%%%%%%%%%%%%%%%%%%%%%%%%%%%%%%%%%%%%%%%%%
\section{Light propagation in plasmic medium}\label{ngse}
%%%%%%%%%%%%%%%%%%%%%%%%%%%%%%%%%%%%%%%%%%%%%%%%%%%%%%%%%%%%%%%%%%

\subsection{{Some backgrounds}}\label{subsec:geometry}

Light propagation in medium is indeed described in the phase space, whose Hamiltonian dynamics gives the structure of the manifold's cotangent bundle. Given the manifold $(\mathcal{M},g_{\alpha\beta})$ expressed in the chart $x^{\alpha}$, the cotangent bundle $T^*\mathcal{M}$ provides the means to define the Hamiltonian 
$H\equiv H(x^\alpha,p_\alpha)$ where $p_\alpha$ is the momentum (wave) covector associated with the cotangent bundle. The Hamilton-Jacobi equation is therefore given in the form 
\begin{equation}\label{eq:H-J}
    H(x^\alpha,p_\alpha) = \frac{1}{2}~ \mathfrak{g}^{\alpha\beta} p_\alpha p_\beta = 0,
\end{equation}
in which $\mathfrak{g}^{\alpha\beta}(x^\alpha)$ is the metric describing $T^*\mathcal{M}$, and is called the \textit{optical metric}. In this sense, the wave (co)vector $p_\alpha$ is considered parallel to the tangential velocity 4-vector $u^\alpha \equiv \dot x^\alpha$\footnote{Here, over-dots indicate $\partial_\tau$ where $\tau$ is the congruence affine parameter.} of the light congruence, i.e. $p_\alpha = \mathfrak{g}_{\alpha\beta} u^\beta$ and according to Eq.~(\ref{eq:H-J}), the light propagates on null congruences with respect to the cotangent bundle. This however is not what an observer on $\mathcal{M}$ would measure, because $p_\alpha\neq g_{\alpha\beta} u^\alpha$ and $g^{\alpha\beta} p_\alpha p_\beta \neq 0$. This means that light behaves like massive particles during its propagation in a medium. In general, such media are given the properties of dielectrics. In fact, the connection between the light propagation in dielectric media and that in the gravitational systems, was recognized in the early days of the advent of general relativity. According to Eddington, relativistic forms of light propagation near a massive object, can be emulated in an appropriate refractive medium \cite{Eddington1920}. In reverse, Gordon pointed out that light propagation in a medium with specific refractive properties, can be emulated in a curved spacetime background endowed with an optical metric inferred from the optical properties of that medium \cite{Gordon1923}. This connection was elaborated further in terms of the effect permittivity ($\varepsilon$) and permeability ($\mu$) of an arbitrary spacetime metric by Plebanski \cite{Plebanski1960} and for the first time, the Gordon's optical metric was used by de Felice to construct (mathematically) a dielectric medium which could mimic a Schwarzschild black hole \cite{deFelice1971}. The Gordon's optical metric is written as \cite{Synge1964}
\begin{equation}\label{eq:opticalmetric}
    \mathfrak{g}^{\alpha\beta} = g^{\alpha\beta} + \left(1-n^2\right) v^\alpha v^\beta,
\end{equation}
where $n(x^\alpha)\equiv\sqrt{\varepsilon\mu}$ and $v^\alpha$ are respectively the scalar refractive index and the tangential velocity 4-vector of the dielectric in the comoving frame\footnote{In fact, since the observer moves on a time-like curve on $\mathcal{M}$, then in the $(- + + +)$ sign convention, $g_{\alpha\beta} v^\alpha v^\beta = -1$. In the same sense, the contraction $v^\alpha p_\alpha$ should be normalized to a real value, which here is the energy of a photon of frequency $\omega$ ($\mathcal{E} = \hbar\omega$), evaluated by an observer, comoving with the plasma.}. In order to include anisotropy, birefringence and magnetoelectric couplings, the notion of the optical metric has been given efforts to be generalized \cite{Ehlers1968,Chen2009a,Chen2009b,Thompson2018}. In the most covariant form, this metric is pseudo-Finslerian, according to the relation
\begin{equation}\label{eq:opticalmetric-covariant}
    \mathfrak{g}^{\alpha\beta} = \frac{\partial^2 H}{\partial p_\alpha \partial p_\beta}.
\end{equation}
In what follows, we consider that a spherically symmetric region (the exterior geometry of a black hole) is filled with a dielectric material, in the form of an inhomogeneous cold plasma with a scalar refractive index. We can therefore assume that the light follows the trajectories on the background described by Gordon's optical metric (\ref{eq:opticalmetric}).

%%%%%%%%%%%%%%%%%%%%%%%%%%%%%%%%%%%
\subsection{{Light propagation in a spherically symmetric plasmic medium surrounding a Weyl black hole}}\label{subsec:lightPropagation}

Weyl gravity is a theory of fourth order in the metric. The simplified form of its action reads as \cite{Mannheim:1989}
\begin{equation}\label{eq:action}
    I_{W}=-2\mathcal{K}\int{\textmd{d}^4x}\sqrt{-g}\,\,\left(R^{\alpha\beta}R_{\alpha\beta}-\frac{1}{3}R^2\right),
\end{equation}
where $\mathcal{K}$ is coupling constant. Applying the principle of least action in the form $\frac{\delta{I_W}}{\delta{g_{\alpha\beta}}} = 0$, leads to the Bach equation $W_{\alpha\beta} = 0$ in which the Bach tensor is defined as \cite{Bach:1921,Szekeres:1968}
\begin{eqnarray}\label{eq:Bach}
W_{\alpha\beta}&=&\nabla^\sigma\nabla_\alpha R_{\beta\sigma}+\nabla^\sigma\nabla_\beta
R_{\alpha\sigma}-\Box R_{\alpha\beta}-g_{\alpha\beta}\nabla_\sigma\nabla_\gamma
R^{\sigma\gamma}\nonumber\\
&&-2R_{\sigma\beta}
{R^\sigma}_\alpha+\frac{1}{2}g_{\alpha\beta}R_{\sigma\gamma}R^{\sigma\gamma}-\frac{1}{3}\Big(2\nabla_\alpha\nabla_\beta
R-2g^{\alpha\beta}\Box R\nonumber\\
&&-2RR_{\alpha\beta}+\frac{1}{2}g_{\alpha\beta}R^2\Big).
\end{eqnarray}
The spherically symmetric vacuum solution to the Bach equation, proposed by Mannheim and Kazanas was in the form \cite{Mannheim:1989}
\begin{equation}\label{eq:generalmetric}
    \ed s^2 = -B(r) \ed t^2 + B(r)^{-1} \ed r^2 + r^2 (\ed\vartheta^2 + \sin^2\vartheta \ed\phi^2),
\end{equation}
where the lapse function $B(r)$ included dark energy and dark matter relevant terms. Here however, we consider a non-vacuum solution for the equation 
\begin{equation}\label{eq:W=T}
   W_{\alpha\beta} = \frac{1}{4\mathcal{K}}~\mathcal{T}_{\alpha\beta},
\end{equation}
in which 
\begin{equation}\label{eq:stress-energy-tensor}
    \mathcal{T}_{\alpha\beta} = M_{\alpha\beta} + \chi_{\alpha\beta},
\end{equation}
is the energy-momentum tensor composed of the massive ($M_{\alpha\beta}$) and the electromagnetic ($\chi_{\alpha\beta}$) parts. Considering the completely static case, in which only the $00$ element of the above tensors takes part, the exterior geometry of a charged black hole on a cosmological background has been obtained as \cite{Payandeh:2012mj}:
\begin{equation}\label{lapse}
	B(r)=1-\frac{r^{2}}{\lambda^{2}}-\frac{Q^2}{4 r^2},
\end{equation}
in which
\begin{equation}\label{par1}
	\frac{1}{\lambda^2}=\frac{3\,\tilde{m}}{\tilde{r}^{3}} +\frac{2\,c_1}{3},\quad Q=\sqrt{2}\,\tilde{q},
\end{equation}
where $\tilde{m}$ and $\tilde{q}$ are respectively the mass and the charge distributed in a source of radius $\tilde{r}$. {Note
that, to reduce this solution to that of the Reissner-Nordstr\"{o}m-(Anti-)de Sitter black hole of mass $M_0$, charge $Q_0$ and cosmological constant $\Lambda$, one is required to consider $\tilde{r}$ as a free radial distance, and do the transformations $3\tilde{m}\rightarrow M_0$, $2 c_1\rightarrow\pm \Lambda$, and $Q\rightarrow 2 i Q_0$, which is an imaginary transformation. Therefore, the reduction of the black hole solutions used in this paper, to the general relativistic black holes, is not trivial.}
It is readily noted that this spacetime allows two horizons; an event horizon $r_+$ together with a cosmological horizon $r_{++}$, placed at
\begin{eqnarray}
	\label{rmas}&&r_+=\lambda\left[ \frac{1}{2}-\sqrt{\frac{1}{4}\left(1-\frac{Q^2}{\lambda^2}
	\right)}\,\right] ^{1/2},\\
		\label{rmm}&&r_{++}=\lambda\left[ \frac{1}{2}+\sqrt{\frac{1}{4}\left(1-\frac{Q^2}{\lambda^2}
	\right)}\,\right] ^{\frac{1}{2}},
\end{eqnarray}respectively. Thus, we can write Eq. (\ref{lapse}) conveniently as
\begin{equation}
    \label{lapse2} B(r)=\frac{(r_{++}^2-r^2)(r^2-r_+^2)}{\lambda^2 r^2}.
\end{equation}
Accordingly, the extremal black hole with a unique horizon at $r_{ex}=r_+=r_{++}=\lambda/\sqrt{2}$ is obtained when $\lambda=Q$ , and the naked singularity appears when $\lambda<Q$.

As mentioned before, we consider that this black hole is surrounded by an inhomogeneous non-magnetized, optically-thin  plasmic shell. The index of refraction of such medium is given by the relation
\begin{equation}
n^2(r)=1-\frac{\omega_p^2(r)}{\omega^2(r)},\label{refraindex}
\end{equation}
where $\omega_p$ is the electron plasma frequency given by
 \begin{equation}\label{elecplasfrec}
\omega_p^2(r)=K_e\,N(r),\quad K_e=\frac{e^2}{\epsilon_0\, m_e}=3182.6\, [{\rm m}^3/{\rm s}^2]. 
\end{equation} Here $N(r)$ is the electron concentration in plasma, $e$ is the electric charge of the electron and $m_e$ is the electron mass. 

For the sake of simplicity, in what follows, we restrict our analysis to the equatorial plane ($\vartheta=\pi/2$); hence, $p_{\vartheta}=0$. Under such condition, applying the optical metric (\ref{eq:opticalmetric}) to the Hamiltonian in Eq. (\ref{eq:H-J}) we get
\begin{eqnarray}\label{hamilt1}
H&=&\frac{1}{2}\left[g^{\alpha\beta}p_\alpha p_\beta+\hbar^2\omega_p^2(r)\right]\nonumber\\
&=&	\frac{1}{2}\left(-\frac{p_t^2}{B(r)}+B(r)p_r^2+\frac{p_{\phi}^2}{r^2}+\hbar^2\omega_p^2(r)\right).
\end{eqnarray}
Accordingly, the canonical Hamilton's equations
\begin{equation}
\label{hamileq}\dot{p}_\alpha=-\frac{\partial H}{\partial x^\alpha}, \qquad \dot{x}^\alpha=\frac{\partial H}{\partial p_\alpha},
\end{equation}
in the cyclic coordinates $(t, \phi)$ yield
\begin{eqnarray}
\label{tpunto}\dot{p}_t&=&-\frac{\partial H}{\partial t}=0\quad \Rightarrow p_t=-\hbar  \omega_0=\textrm{cte.},\\
\label{phipunto}\dot{p}_{\phi}&=&-\frac{\partial H}{\partial \phi}=0\quad \Rightarrow p_{\phi}=\ell=\textrm{cte.},
\end{eqnarray}
regarding which, we can infer that $\hbar \omega_0  \equiv\mathcal{E}_0$ and $\ell$ are constants of motion, associated with its temporal and rotational invariance. The remaining equations read
\begin{eqnarray}\nonumber
\dot{p}_r&=&-\frac{\partial H}{\partial r}=\\\label{rpunto}
&=&\frac{\ell^2}{r^3}-\frac{{\rm d}}{{\rm d}r}\left[\frac{\hbar^2 \omega_p^2(r)}{2}\right]-\frac{1}{2}\frac{{\rm d}B(r)}{{\rm d}r}\left[p_r^2+\frac{\hbar^2 \omega_0^2}{B^2(r)}\right],\\
\label{tdot}\dot{t}&=&\frac{\partial H}{\partial p_t}=-\frac{p_t}{B(r)}=\frac{\hbar \omega_0}{B(r)},\\
\label{phidot}\dot{\phi}&=&\frac{\partial H}{\partial p_{\phi}}=\frac{p_{\phi}}{r^2}=\frac{\ell}{r^2},\\
\label{rdot}\dot{r}&=&\frac{\partial H}{\partial p_r}=B(r)\, p_r.
\end{eqnarray}
There is also one extra condition 
\begin{equation}
\label{hcero}0=\frac{\ell^2}{r^2}+B(r)\, p_r^2-\left[\left(\frac{\hbar \omega_0}{\sqrt{B(r)}}\right)^2-\hbar^2 \omega^2_p(r)\right],
\end{equation}
inferred from the Hamilton-Jacobi equation. Note that, the radial dependence of the photon's frequency, measured by the comoving  observer, is obtained by the redshift formula 
\begin{equation}\label{redshgrav}
\omega(r)=\frac{\omega_0}{\sqrt{B(r)}}.
\end{equation}
Therefore, it is no hard to see from Eqs. (\ref{hcero}) and (\ref{redshgrav}) that, in a given position $r$, the photon frequency $\omega(r)$ is bigger than the plasma frequency $\omega_p(r)$, i.e. 
\begin{equation}\label{c1}\omega(r)>\omega_p(r),
\end{equation} which is an empirical constraint for light propagation in plasma \cite{Perlick:2015vta}.

Now, turning to the subject in hand, we commence studying the light propagation in the above system. Using Eqs. (\ref{phidot}) and (\ref{rdot}), the general orbits are governed by
\begin{equation}
\label{eqmot}\left(\frac{{\rm d}r}{{\rm d}\phi}\right)^2=\frac{\dot{r}^2}{\dot{\phi}^2}=\mathfrak{F}(r),
\end{equation}
with 
\begin{equation}
\label{ffunct}\mathfrak{F}(r)=r^2 B(r)\left[\frac{h^2(r)}{h^2(\mathcal{R})}-1\right],
\end{equation}
where
\begin{eqnarray}
\label{hfunct}h^2(r)&=&\frac{r^2\,n^2(r)}{B(r)}=\frac{r^2}{B(r)}\left(1-\frac{\omega_p^2(r)}{\omega^2(r)}\right),\\ \label{hRfunct}h^2(\mathcal{R})&=&\frac{\ell^2}{\hbar^2 \omega_0^2}=b^2.
\end{eqnarray}
Equation (\ref{hRfunct}) relates the closest approach to the source, $\mathcal{R}$, to the impact parameter $b$ (another constant of motion). Therefore, exploiting Eqs.~(\ref{eqmot}) to (\ref{hRfunct}), the deflection angle for a light ray traveling from $r_{++}$ to $\mathcal{R}$ and goes again to $r_{++}$, can be calculated as
\begin{eqnarray}\nonumber
\widehat{\alpha}&=&2 b\int_{R}^{r_{++}}\frac{{\rm d} r}{\sqrt{r^2 B(r) h^2(r)-b^2 r^2 B(r)}}-\pi\\ \label{defl}
&=&2 b\int_{R}^{r_{++}}\frac{{\rm d} r}{\sqrt{r^4 n^2(r)-b^2 r^2 B(r)}}-\pi.
\end{eqnarray}
The above deflection, relates to the lensing effect caused by massive sources. This shows that how outermost objects can change their apparent position. The above deflection angle however could be determined specifically, once $n(r)$ is given an appropriate algebraic expression, regarding the causality conditions. We deal with such expressions in the next section.

%%%%%%%%%%%%%%%%%%%%%%%%%%%%%%%%%%%%%%%%%%%%%%%%%%%%%%
\section{Specific cases of $n(r)$}\label{sec:nAnzats}

\begin{figure}[t]
    \centering
    \includegraphics[width=9cm]{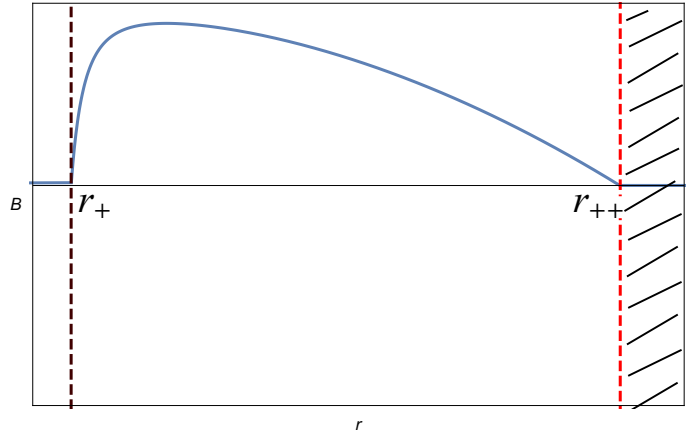}
    \caption{The causal structure offered by a charged Weyl black hole. Events outside $r_{++}$  does not have casual connections with the observers residing inside it.}
    \label{fig:causal}
\end{figure}

The casual connection in the spacetime constructed by the charged Weyl black hole in Eq.~(\ref{lapse}), suggests that an observer inside the cosmological horizon cannot be aware of the events from the region covered by $r>r_{++}$ (see Fig.~\ref{fig:causal}). For this reason, any algebraic assignment for the refractive index $n(r)$ should respect this kind of causality. This means that the refraction is well-defined only inside the boarders of the casual connection.   
Accordingly, we propose relevant algebraic forms, regarding the boundaries of the causality.

%%%%%%%%%
\subsection{First ansatz}\label{subsec:ansatz1}
\begin{figure}[t]
    \centering
    \includegraphics[width=9cm]{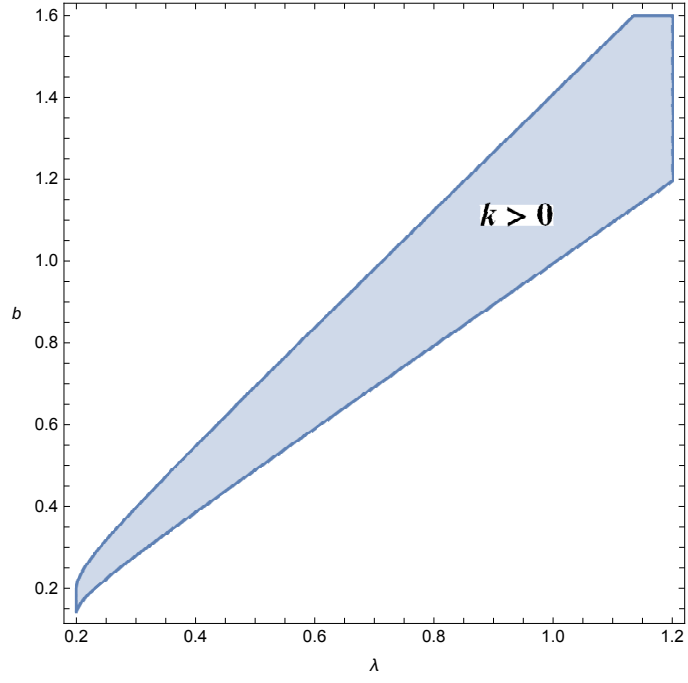}
    \caption{The region of allowed values for $b$ for which the condition $k>0$ is satisfied. The plot has been done for $Q=0.1$. The considered range for $b$ is from $1.02\, r_{++}$ to $1.4\, r_{++}$ for the given $\lambda$ and $Q$, so that it can cover the allowed values.}
    \label{fig:delta1-region}
\end{figure}

\begin{figure}[t]
    \centering
    \includegraphics[width=10cm]{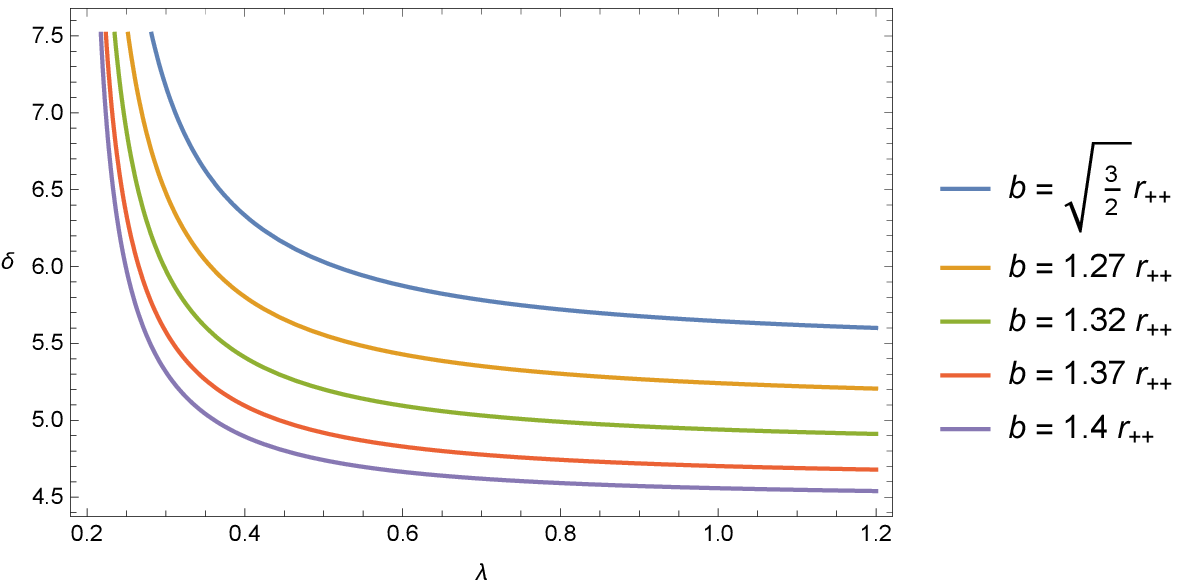}~(a)
     \includegraphics[width=10cm]{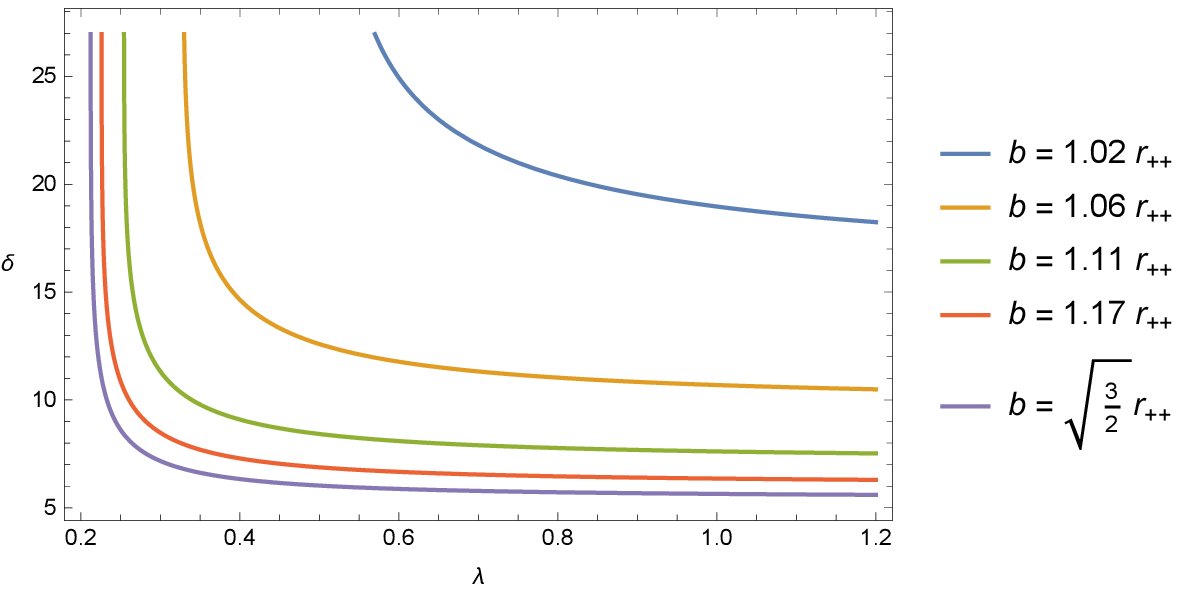}~(b)
    \caption{The range for the deflection angles obtained from the refractive index of the first kind, for five given values for the impact parameter, within the allowed range for each case. We have taken $Q=0.1$ and the plots have been done for (a) $\sqrt{\frac{3}{2}}\,r_{++}\leq b < \sqrt{2}\, r_{++}$ and (b) $r_{++}<b\leq\sqrt{\frac{3}{2}}\,r_{++}$. As it is seen, the second condition makes the deflection to change more rapidly toward the stable value.}
    \label{fig:delta1}
\end{figure}

Taking into account a case in which $n(r_{++}) = n(r_+) = 0$, we propose the following ansatz:
\begin{equation}\label{eq:ansatz1}
    n^2(r) = B(r)\left[
    \frac{r_{++}^2}{r^2}+1
        \right],
\end{equation}
which of course, has its maximum at $r_+<r_{\max}<r_{++}$. By means of Eq.~(\ref{lapse2}), this can be rewritten as
\begin{equation}\label{eq:ansatz1-recast}
    n^2(r)=\frac{(r_{++}^4-r^4)(r^2-r_+^2)}{\lambda^2 r^4}.
\end{equation}
The integrand in Eq.~(\ref{defl}) is $\frac{1}{\sqrt{\mathfrak{P}(r)}}$, in which, according to the above definition, we have
\begin{equation}\label{eq:p(r)-1}
    \mathfrak{P}(r) = \frac{(r_{++}^2-r^2)(r^2-r_+^2)}{\lambda^2}(r^2-\mathcal{R}^2).
\end{equation}
Here, $\mathcal{R} = \sqrt{b^2-r_{++}^2}$ is the closest approach as appeared in Eq.~(\ref{hRfunct}). This implies that $b>r_{++}$. Now, recasting 
\begin{equation}\label{eq:p(r)-2}
\mathfrak{P}(r) = \frac{r^6 r_{++}^2 r_+^2 \mathcal{R}^2}{\lambda^2}\left(
\frac{1}{r^2}-\frac{1}{r_{++}^2}
\right)
\left(
\frac{1}{r_+^2}-\frac{1}{r^2}
\right)
\left(
\frac{1}{\mathcal{R}^2}-\frac{1}{r^2}
\right),
\end{equation}
we can rewrite the deflection angle in Eq.~(\ref{defl}) as
\begin{eqnarray}\label{eq:delta1}
    \delta &\equiv& \widehat{\alpha} + \pi\nonumber\\
    &=&\frac{b \lambda}{r_{++} r_+ \mathcal{R}} \int_{0}^{\xi_{++}}\frac{\ed \xi}{\sqrt{\xi (\xi_{++}-\xi) (\xi_+ + \xi)}},
\end{eqnarray}
for which, we have used the change of variable
\begin{equation}\label{eq:rtoxi}
    \xi(r) \doteq \frac{1}{\mathcal{R}^2} - \frac{1}{r^2},
\end{equation}
and have defined
\begin{eqnarray}\label{eq:xi++}
\xi_{++} &\doteq& \xi(r_{++}),\\
\xi_+ &\doteq& \frac{1}{r_+^2}-\frac{1}{\mathcal{R}^2}. 
\end{eqnarray}
The integral in Eq.~(\ref{eq:delta1}) is in fact an elliptic integral of the first kind. We therefore get 
\begin{equation}\label{eq:delta2}
    \delta = \frac{b \lambda}{r_{++} r_+ \mathcal{R}} \,\bar{\bar{g}} \,K(k)
\end{equation}
in which \cite{handbookElliptic}
\begin{eqnarray}\label{eq:ellipticDefinitions}
\bar{\bar{g}} &=& \frac{2}{\sqrt{\xi_{++}+\xi_+}}= \frac{2 r_{++} r_+}{\sqrt{r_{++}^2 - r_+^2}},\\
K(k) &\equiv& F\left(\varphi(\xi_{++}),k\right) \nonumber\\
  &=& \int_0^{\frac{\pi}{2}} \frac{\ed\eta}{\sqrt{1-k^2 \sin^2\eta}},
\end{eqnarray}
where the latter is the complete elliptic integral of the first kind, given
\begin{eqnarray}\label{eq:varphi(y)}
   \varphi(y) &=& \arcsin\left(
   \sqrt{\frac{\xi_{++}+\xi_{+}}{\xi_{++}}\,\frac{y}{y+\xi_{+}}}
   \right),\\
   k &=& \sqrt{\frac{\xi_{++}}{\xi_{++}+\xi_{+}}} = \frac{r_{+}}{\mathcal{R}}\sqrt{\frac{r_{++}^2-\mathcal{R}^2}{r_{++}^2-r_+^2}}\label{eq:k}.
\end{eqnarray}
Regarding the relation between $b$ and $\mathcal{R}$, the deflection could be rewritten in terms of either of the above parameters as
\begin{subequations}
\begin{align}
\delta(\lambda,b) = \frac{2 b \lambda}{\sqrt{(b^2-r_{++}^2)(r_{++}^2-r_+^2)}}~  K\left(k(b)\right),\\
\delta(\lambda,\mathcal{R}) = \frac{2 \lambda}{\mathcal{R}}\, \sqrt{\frac{\mathcal{R}^2+r_{++}^2}{r_{++}^2-r_+^2}}~  K\left(k(\mathcal{R})\right).
\end{align}
\end{subequations}
Note that, not all values of $b$ are allowed for the light ray trajectories. Since $b > r_{++}$ and $k>0$, regarding Eq.~(\ref{eq:k}), we have either $\sqrt{\frac{3}{2}}\,r_{++}\leq b < \sqrt{2}\, r_{++}$ or $r_{++}<b\leq\sqrt{\frac{3}{2}}\,r_{++}$. This has been shown in Fig.~\ref{fig:delta1-region}. 
Also, the behavior of $\delta$ has been demonstrated in Fig.~\ref{fig:delta1}, distinctly for the above two categories. 
The plots show that the second kind of confinement for $b$, results in more fast varying deflections.

%%%%%%%%%%%%%%%%%%%%%%%%%%%%%%%%%%%%%%%%%%%%%%
\subsection{Second ansatz}\label{subsec:ansatz2}

\begin{figure}[t]
    \centering
    \includegraphics[width=9.5cm]{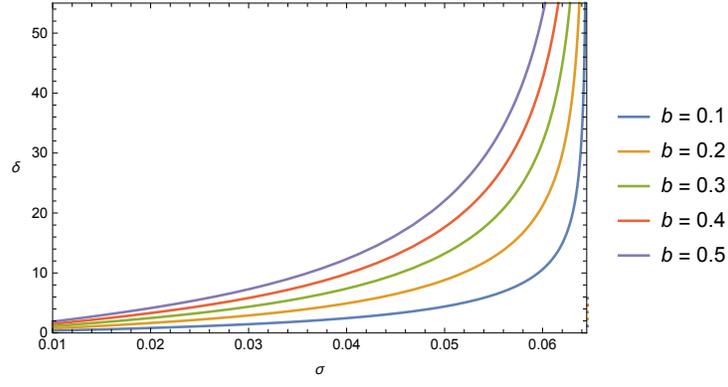}
    \caption{The behavior of the deflection angle obtained from the refractive index of the second kind, for five different impact parameters. The smaller the impact parameter is, the faster $\delta$ increases. The asymptotic behavior however stems in the presence of elliptic integrals in the description of $\delta$. In this figure, we can see that for a certain value of $\sigma$, the light rays escape from the black hole. 
    The plots have been done for $Q=0.1$ and $\lambda = 0.25$ (in arbitrary length units).}
    \label{fig:delta2}
\end{figure}

As the second guess, we consider a more complicated algebraic form, reading
\begin{equation}\label{eq:ansatz2}
    n^2(r) = \frac{B(r)}{r^2}\left[
    b^2+(r^2+\sigma^2)^2\left(r^2-r_{++}^2\left(1-\frac{\sigma^2}{r_+^2}\right)\right)
    \right],
\end{equation}
in which $\sigma\equiv\sigma(r_+,r_{++})$ is a function whose value satisfies the condition $0<\sigma < r_+$. Exploiting this in the integrand, we get
\begin{equation}\label{eq:p(r)new-1}
\mathfrak{P}(r) = \frac{1}{\lambda^2}\left[
 (r^2-r_+^2)(r_{++}^2-r^2)(r^2+\sigma^2)^2(r^2-\mathcal{R}^2)
\right],
\end{equation}
where the newly defined closest approach is $\mathcal{R}=\sqrt{r_{++}^2\left(1-\frac{\sigma^2}{r_+^2}\right)}$. Upon recasting, the above polynomial becomes
\begin{eqnarray}\label{eq:p(r)new-2}
    \mathfrak{P}(r) &=& \left(
    \frac{r^5 r_+ r_{++} \sigma^2 \mathcal R^2}{\lambda}
    \right)\left(
    \frac{1}{r_+^2}-\frac{1}{r^2}
    \right)\left(
    \frac{1}{r^2}-\frac{1}{r_{++}^2}
    \right)\left(
    \frac{1}{r^2}+\frac{1}{\sigma^2}
    \right)^2\left(
    \frac{1}{\mathcal R^2}-\frac{1}{r^2}
    \right).
\end{eqnarray}
Applying the same change of variable as in Eq. \eqref{eq:rtoxi}, we get 
\begin{equation}\label{eq:delta-new}
    \delta = \frac{b \lambda }{ r_+ r_{++} \mathcal{R} \sigma} 
   \mathcal{I}_1,
\end{equation}
where
\begin{equation}\label{eq:intergalI}
    \mathcal{I}_1=\int_{0}^{\xi_{++}} \frac{\left(
    \xi-\frac{1}{\mathcal{R}^2}     
    \right)~\mathrm{d}\xi}{(\xi-\bar\xi)\sqrt{\xi (\xi_++\xi) (\xi_{++}-\xi)}}.
\end{equation}
Here we have defined
\begin{equation}\label{eq:xibar}
    \bar\xi \doteq \frac{1}{\mathcal{R}^2}+\frac{1}{\sigma^2},
 \end{equation}
and other definitions remain the same as in the previous case. The integral in Eq.~\eqref{eq:intergalI} has an elliptic counterpart so that we can rewrite it as \cite{handbookElliptic}
\begin{equation}\label{eq:intergalI-1}
    \mathcal{I}_1 = \frac{\bar{\bar{g}}}{\mathcal{R}^2 \bar\xi}
\int_0^{K(k)}\frac{1-\beta_1^2~\mathrm{sn}^2(\eta)}{1-\beta^2 \mathrm{sn}^2(\eta)}\mathrm{d}\eta,
\end{equation}
in which 
\begin{equation}\label{eq:betas-in-xi}
       \beta^2 = \frac{\frac{1}{\mathcal{R}^2}(\xi_+ + \bar\xi)}{\bar\xi (\xi_{+}+\frac{1}{\mathcal{R}^2})}\,\beta_1^2
       = \frac{\xi_{++} (\xi_+ + \bar\xi)}{\bar\xi (\xi_{++}+\xi_+)},
\end{equation}
 and $\mathrm{sn}(\eta)$ is a Jacobi elliptic function, doubly periodic in $\eta$, and is defined as \cite{handbookElliptic}
\begin{equation}\label{eq:jacobiSN}
    \mathrm{sn}(\eta) = \sin(\varphi), 
\end{equation}
with $\varphi$ given in Eq.~\eqref{eq:varphi(y)}. Considering the above elliptic counterpart, we get
\begin{equation}\label{eq:intergalI-2}
    \mathcal{I}_1 = \frac{\bar{\bar{g}}}{\mathcal{R}^2 \beta^2 \bar{\mu}}\left[
    \beta_1^2 K(k) + (\beta^2-\beta_1^2) \Pi(\beta^2,k)
    \right],
\end{equation}
where 
\begin{equation}\label{eq:Pi(beta)}
    \Pi(\beta^2,k) = \int_0^{\frac{\pi}{2}} \frac{\mathrm{d}\eta}{(1-\beta^2 \sin^2\eta)\,\sqrt{1-k^2 \sin^2\eta}}
\end{equation}
is the complete elliptic integral of the third kind. With this in mind, and taking into account the definition in Eq.~\eqref{eq:ellipticDefinitions}, we finally get 
\begin{equation}\label{eq:delta-new-4}
    \delta = \frac{2 b \lambda}{\mathcal{R} \sigma \left(
    1+\frac{\mathcal{R}^2}{\sigma^2}
    \right)\,\sqrt{r_{++}^2-r_+^2}}
    \left(
    \frac{r_+^2+\sigma^2}{\mathcal{R}^2+\sigma^2}\,K(k)
    +\frac{\mathcal{R}^2-r_+^2}{\mathcal{R}^2+\sigma^2}
    \,\Pi(\beta^2,k)
    \right),
\end{equation}
which is compatible with
\begin{equation}\label{eq:betas-in-r}
       \beta^2 = \frac{\mathcal{R}^2 +\sigma^2}{r_+^2 + \sigma^2}\beta_1^2 = \frac{(r_{++}^2-\mathcal{R}^2) (r_+^2 + \sigma^2)}{(r_{++}^2-r_+^2) (\mathcal{R}^2+\sigma^2)},
       \end{equation}
and $k = (r_+/\mathcal{R})\beta_1$. Note that, since $b$ does not have any contribution in the parameter $\mathcal{R}$, this angle does not put any restrictions on the impact parameter and the condition $k>0$ is always satisfied. The behavior of the deflection in Eq.~\eqref{eq:delta-new-4} has been plotted in Fig.~\ref{fig:delta2} for some different impact parameter. The asymptotic behavior of the plots, stems in the elliptic functions included in the description of $\delta$. Similar behavior was observed in Fig.~\ref{fig:delta1}. Physically, this means that light rays with definite impact parameters, can only contribute to the lensing process of black holes with definite physical properties (namely $\lambda$ and $Q$). So, for certain black holes, not all rays can provide imaging through gravitational lensing. In the plots of Fig.~\ref{fig:delta2}, light ray deflections are given in terms of changes of the parameter $\sigma$. \\

In this section, we talked about two completely different possibilities of the radial dependence of the refractive index. This parameter tells us about how light can deviate during its travel inside the plasma and in our case, at the same time, how can be affected by the background geometry. The obtained deflection angles, corresponding to these specific cases of the refractive index, demonstrate the ability of the plasma to contribute in the usual spacetime curvature caused by the black hoe. However, once the deflection is so high, in a way that the light rays are confined to circulating on a surface around a black hole, they form a photon surface which constitutes the foundations of the so-called black hole shadow. In the next section, we exploit the recently assessed forms of $n^2(r)$ to investigate the characteristics of the corresponding photon surfaces.

%%%%%%%%%%%%%%%%%%%%%%%%%%%%%%%%%%%%%%%%%%%%%%%%%%%%%
\section{The photon sphere}\label{sec:rph}

Photon spheres are those hypersurfaces, on which light rays can stay on a stable circular path. The innermost photon sphere has the radius $\mathcal{R}$ introduced above. The photon surfaces however can be determined by analyzing purely angular light orbits. This condition requires $\dot r = \ddot r =0$, that from Eq.~(\ref{rdot}) it follows that $p_r = 0$. We therefore can rewrite the Hamilton-Jacobi equation as
\begin{equation}\label{eq:ell2-1}
    \ell^2 = \hbar^2 r^2 \left[
    \frac{\omega_0^2}{B(r)} - \omega_p^2(r)
    \right].
\end{equation}
Furthermore, differentiating Eq.~(\ref{rdot}) with respect to the affine parameter, results in 
\begin{equation}\label{eq:pdotr-new}
    \dot{p}_r = \frac{1}{B(r)}\left(\ddot r - \frac{\ed B(r)}{\ed r} \dot r p_r\right),
\end{equation}
according to which, the zero radial velocity condition implies $\dot p_r = 0$. Hence, Eq.~(\ref{rpunto}) can be recast as
\begin{equation}\label{eq:ell2-2}
    \ell^2 = \frac{\hbar r^3}{2} \left[
    \frac{\ed \omega_p^2(r)}{\ed r}
    +\frac{\ed B(r)}{\ed r}\left(
    \frac{\omega_0^2}{B^2(r)}
    \right)
    \right].
\end{equation}
Subtracting the above equations and after some manipulations, we get the equation governing the radius of the circular light orbits
\begin{equation}\label{eq:lightOrbit-1}
    \frac{\ed}{\ed r} h^2(r) = 0.
\end{equation}
Solutions to this equation determine the radius of photon spheres. Satisfaction of Eq.~\eqref{eq:lightOrbit-1} is done by letting $h^2(r)=\mathfrak{c}=\mathrm{const.}$ Applying this in Eq.~\eqref{hfunct} and taking into account the redshift in Eq.~\eqref{redshgrav} we get
\begin{equation}\label{eq:omegap-new-1}
    \omega_p^2(r) = \frac{\omega_0^2}{B(r)}\left(
    1-\frac{\mathfrak{c}\,B(r)}{r^2}
    \right).
\end{equation}
This demands the following condition 
{for $r>r_+$}: 
\begin{equation}\label{eq:condition-c}
    \frac{r^2}{B(r)}>\mathfrak{c}.
\end{equation}
Furthermore, considering Eq.~\eqref{refraindex} in Eq.~\eqref{eq:lightOrbit-1} we get
\begin{eqnarray}\label{eq:lightOrbit-2}
        0 &=& \left(2 B(r) - r \left(\frac{\mathrm{d}}{\mathrm{d} r} B(r)\right)\right)\left(
    1-B(r)\frac{\omega_p^2(r)}{\omega_0^2}
    \right)\nonumber\\
  &&  - r\, B(r) \left[
    \left(\frac{\mathrm{d}}{\mathrm{d} r} B(r)\right)\frac{\omega_p^2(r)}{\omega_0^2} + \frac{2 B(r) \omega_p(r)}{\omega_0^2} \left(\frac{\mathrm{d}}{\mathrm{d} r} \omega_p(r)\right)
    \right].
\end{eqnarray}
In the case of no plasmic surroundings, we have $\omega_p(r) = 0$, yielding the following photon sphere radius in vacuum: \begin{equation}\label{eq:rphoton-vacuum}
    r^{(\mathrm{vac})}_\mathrm{ph} = \frac{\sqrt{2}\, r_+\,r_{++}}{\sqrt{r_+^2+r_{++}^2}}.
\end{equation}
From the values in Eqs.~\eqref{rmas} and \eqref{rmm}, this gives $r^{(\mathrm{vac})}_\mathrm{ph} = Q/\sqrt{2}$, which is the same as the radius of the critical orbits, $r_c$, obtained in Ref.~\cite{Fathi:2020} for the same black hole in vacuum\footnote{Note that, the radius in Eq.~\eqref{eq:rphoton-vacuum} will never regain the famous Schwarzschild $r=3 M$ photon sphere, by letting $r_+=r_{++} = 2 M$. This is because the metric potential in Eq.~\eqref{lapse} is totally different in structure, regarding the presence and the definition of the $\lambda$ parameter.}.

However, in the presence of plasma, this photon sphere is characterized by solving Eq.~\eqref{eq:lightOrbit-2}. Considering Eq.~\eqref{lapse2}, this differential equation yields
\begin{equation}\label{eq:omegap-new-2}
    \omega_p^2(r) = \frac{\lambda^2 \omega_0^2\left(
    r^2 (r_+^2+r_{++}^2)-r_+^2 r_{++}^2
    \right)}{r^2 (r^2-r_+^2) (r_{++}^2 - r^2)}.
\end{equation}
Note that, as long as the condition $\lambda> Q$ is satisfied, the positivity of the right hand side of the above relation is guaranteed. 

\begin{figure}[t]
    \centering
    \includegraphics[width=10cm]{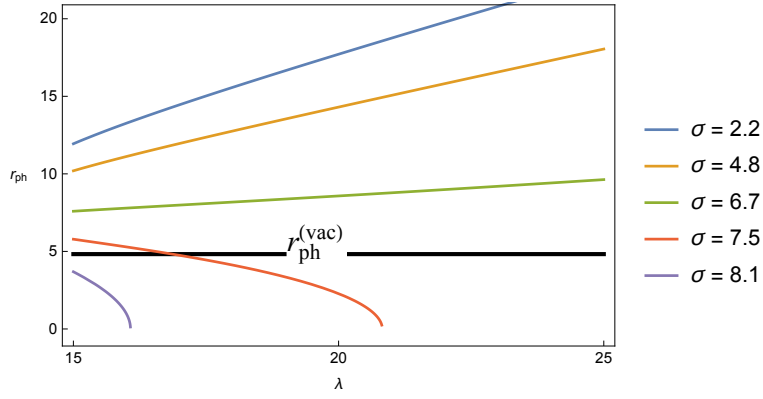}
    \caption{Confronting the radii of vacuum and plasmic photon spheres. The plots have been done for $b = 10$, $Q = 7$ and five different values of $\sigma$ which have been selected according to $\sigma<r_+$. Changes in $b$ do not have any effects on the form of the curves. Obviously, the value of $r_{\mathrm{ph}}^{\mathrm{(vac)}}$ does not depend on $\lambda$ and is therefore a constant in this regard. This is while the plasmic $r_{\mathrm{ph}}$ raises constantly for smaller values of $\sigma$, whereas it drops fast for larger ones. }
    \label{fig:rph}
\end{figure}

Given the frequency in Eq.~\eqref{eq:omegap-new-2}, the radius $r_{\mathrm{ph}}$ now depends on one other characteristic of the plasmic medium, namely the refractive index. This can be seen through Eq.~\eqref{refraindex}, providing $\omega_p^2(r) = (\omega_0^2/B(r))(1-n^2(r))$. This, together with Eq.~\eqref{eq:omegap-new-2}, results in the following alternative for the refractive index:
\begin{equation}\label{eq:n2(r)-new}
    n^2(r) = 1-\frac{(r_+^2+r_{++}^2)}{r^2}+\left(
    \frac{r_+ r_{++}}{r^2}
    \right)^2.
\end{equation}
The determination of $r_{\mathrm{ph}}$ however, requires other definitions for $n^2(r)$. To deal with this, we therefore recall the specific cases discussed in the previous section.
\begin{itemize}
    \item{For the first ansatz in Eq.~\eqref{eq:ansatz1-recast} (plasma of the first kind (PFK)), Eq.~\eqref{eq:n2(r)-new} provides $r_{\mathrm{ph}} = r_+$. {This means that the corresponding hypersurface, formed as the 3-dimensional (3D) closure of the 2D circles characterized by $r=r_{+}$, is indeed a null surface. Although this result could seem unexpected, we here refer the reader to the fact that this photon surface is observed through a dispersive medium (plasma) that based on the geometric structure of the respected refractive index, could affect the photon surface to be located differently from that in the vacuum.}}
    
    \item{For the case in  Eq.~\eqref{eq:ansatz2} (plasma of the second kind (PSK)), we get
    \begin{eqnarray}\label{eq:rph-secondn(r)}
        r_{\mathrm{ph}} &=& \frac{\mathcal{A}^{-\frac{1}{6}}}{\sqrt{6}\,r_+} \left[
        2^\frac{4}{3}(r_+ r_{++})^4+2(r_+ r_{++})^2 \mathcal{A}^\frac{1}{3}
        +(-2\mathcal{A})^\frac{2}{3}\right.\nonumber\\
        &&\left.+\sigma^2\left(
        2^\frac{7}{3}\left(r_+ r_{++}\right)^2\left(r_+^2-r_{++}^2\right)
        -2\left(2r_+^2+r_{++}^2\right)\mathcal{A}^\frac{1}{3}
        \right)\right.\nonumber\\
        &&\left.+\sigma^4\left(
        2^\frac{4}{3}(r_+^4+r_{++}^4)-2^\frac{7}{3}(r_+ r_{++})^2
        \right)
        \right]^{\frac{1}{2}},
    \end{eqnarray}
    with
    \begin{equation}\label{eq:A}
        \mathcal{A}=\sqrt{\mathcal{B}^2-4\left(
        (\sigma\, r_{++})^2+r_+^2(r_{++}^2+\sigma^2)
        \right)^6}-\mathcal{B},
    \end{equation}
    where
    \begin{eqnarray}\label{eq:B}
        \mathcal{B} &=& 27 b^2 r_+^6 + 2 (\sigma r_{++})^6 
           + 6\, \sigma^4 \left(
        r_{++}^2+\sigma^2
        \right) (r_+ r_{++})^2\left[\left(
        r_{++}^2+\sigma^2
        \right)\,r_+^2-r_{++}^2\right]\\
       &&-r_+^6\left(
       2r_{++}^6-27\lambda^2+6\,\sigma^2r_{++}^2(r_{++}^2+\sigma^2)+2\sigma^6
       \right).
       \end{eqnarray}
     }
\end{itemize}
In Fig.~\ref{fig:rph}, we have confronted the above radius for different values of $\sigma$, with the radius of the photon sphere in the vacuum case. We have considered a fixed $b$, because the curves with different values of $b$ will coincide. The vacuum photon sphere exhibits a constant size, whereas the plasmic one can change its radius, depending on the value of $\sigma$. It is observed that, increase in $\lambda$ has different effects on $r_{\mathrm{ph}}$, depending on the corresponding $\sigma$. This means that, the small--$\sigma$ photon spheres expand as $\lambda$ increases, whereas the large--$\sigma$ ones would shrink.\\

In this section, the light rays were considered to travel on a circular path around the black hole and we discussed the outcome of the combination of the background geometry and plasma in confining a photon sphere. This sphere defines the boundary of the black hole's shadow. Now, before going any further on this, let us examine the refractive plasmas under study, in a more physical context.

%%%%%%%%%%%%%%%%%%%%%%%%%%%%%%%%%%%%%%%%%%%%
\section{The implications for $N(r)$}\label{sec:Nr}

Even though the spacetime effects are imposed on the description of the refractive index, nevertheless, the physical interpretation of the particle distribution inside the spacetime is given by the concentration function $N(r)$. Applying the definition given in Eqs.~\eqref{refraindex} and \eqref{elecplasfrec}, we get
\begin{equation}\label{eq:N(r)-definition}
    N(r) = \frac{\omega_0^2}{K_e B(r)}\left(1-n^2(r)\right).
\end{equation}
In this section, paying attention to this quantity we go deeper into the physical implications of both kind plasmas.\\

\begin{figure}[t]
    \centering
    \includegraphics[width=10cm]{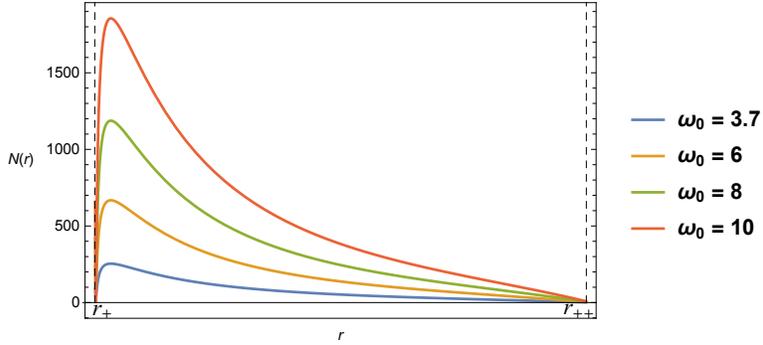}
    \caption{The behavior of particle concentration $N_1(r)$ for four values of $\omega_0$, in the region between the horizons. All four concentrations have a maximum at the same radial distance and at the horizons, $N_1$ is indefinite. It however tends to zero at the vicinity of both horizons. The plots have been done for $Q=7$, $\lambda=19.4$, $b=9$ and we have absorbed $K_e$ into $\omega_0$ (all values are in arbitrary length units). }
    \label{fig:N1(r)}
\end{figure}

The PFK generates
\begin{equation}
N_1(r) = \frac{\omega_0^2\left[r^4\lambda^2-\left(r^2-r_+^2\right)\left(r_{++}^4-r^4\right)\right]}{r^2 \left(
r^2-r_+^2
\right)
\left(
r_{++}^2-r^2
\right)},
\end{equation}
the behavior of which has been illustrated in Fig.~\ref{fig:N1(r)}
inside the causal region. For the PSK,
\begin{figure}[t]
    \centering
    \includegraphics[width=11cm]{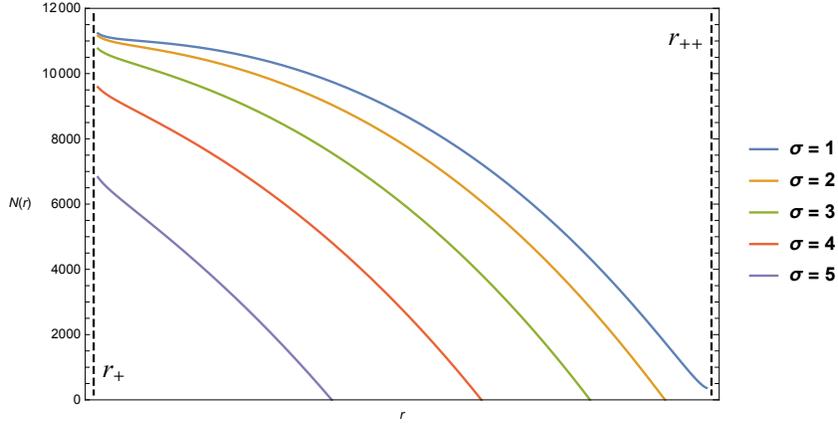}
    \caption{The evolution of particle concentration $N_2(r)$ for five values of $\sigma$, in the region between the horizons. The concentration drops by moving toward $r_{++}$. Significantly, larger $\sigma$ results in a less steep decrease in the concentration. The plots have been done for $Q=7$, $\lambda=15$, $b=10$ and $\omega_0 = 1.37$ (we have absorbed $K_e$ into $\omega_0$ and all values are in arbitrary length units). The above values have been chosen to obtain a good scale of observation and alternations in these values will just change the scale of the plots, not their form of behavior. }
    \label{fig:N(r)}
\end{figure}
the concentration becomes
\begin{equation}\label{eq:N(r)-2}
    N_2(r) = \frac{\omega_0^2}{K_e r^2}\left[
    \frac{r^4 \lambda^2}{(r^2-r_+^2)(r_{++}^2-r^2)}-b^2-(r^2+\sigma^2)^2\left(
    r^2-r_{++}^2\left(
   1-\frac{\sigma^2}{r_+^2}
    \right)
    \right)
    \right],
\end{equation}
which evolves as plotted in Fig.~\ref{fig:N(r)} for five different values of $\sigma$, in the region $r_+<r<r_{++}$. As it is expected, the concentration drops from its highest values at the vicinity of $r_+$, by moving toward $r_{++}$. As we can see from the plots of $N_1(r)$ and $N_2(r)$ (for definite values of $\sigma$), the electron concentration can tend to zero long before reaching the cosmological horizon (where the concentration should be indefinite). One important implication of this property, is that the effect of the plasma can be seen in regions outside its presence, because the refraction $n(r)$ is available in all the region $r_+<r<r_{++}$. This can be interpreted as a combination of electromagnetic effects and optical gravity, manifesting themselves through the refractive index. 
For the second kind plasma, the fall in the value of $N_2(r)$ happens faster for smaller $\sigma$. However we should bear in mind that, through their relation to the horizons, every pair $(\lambda,Q)$ is related to a range for $\sigma$, which has to satisfy $0<\sigma<r_+$.\\ 

\begin{figure}[t]
    \centering
    \includegraphics[width=10cm]{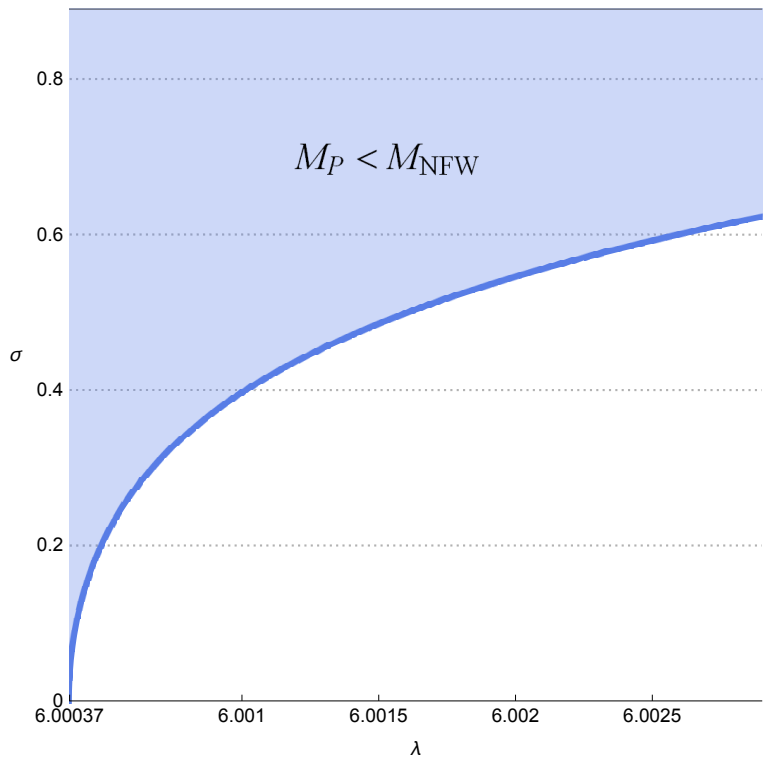}
    \caption{The numerical evaluation of the $M_P \leq N_{\mathrm{NFW}}$ condition. The plot has been done for $b=5.2$, $Q=3$, $\omega_0=2$, $\rho_0 = 1$, $\mathfrak{r}_s = 0.6$ and $\mathfrak{r}_{\mathrm{max}} = 10$. The solid blue line shows the possibility of having a plasmic electron distribution, which obeys the NFW cold dark matter density profile. }
    \label{fig:MOvsMNFW}
\end{figure}

As a matter of interest, let us think of the PSK as a spherically symmetric halo, filling the region $r_+<r<r_{++}$. Although electrons are not usually considered as dark matter candidates, however, it may be of interest to revisit their plasmic distribution in the cold dark matter realm.  In this regard, we therefore compare the total masses obtained from the above particle concentration, and that given by the Navarro-Frenk-White (NFW) density profile. The NFW profile for a cold dark matter distribution is \cite{Navarro1995c,Navarro1996}  
\begin{equation}\label{eq:NFWprofile}
    \rho(r) = \frac{\rho_0}{\frac{r}{\mathfrak{r}_s}\left(
    1+\frac{r}{\mathfrak{r}_s}
    \right)^2},
\end{equation}
in which the initial density $\rho_0$ and the scale radius $\mathfrak{r}_s$ depend on the characteristics of the halos. The integrated mass of the halo is obtained by integrating the above profile within the total volume. Considering a spherically symmetric halo, we obtain
\begin{eqnarray}\label{eq:Mnfw}
    M_{\mathrm{NFW}} &=& \int_0^{\mathfrak{r}_{\mathrm{max}}}\rho(r)\,4\pi r^2  \ed r\nonumber\\
    & =& 4\pi\rho_0 \mathfrak{r}_s^3\left[
    \ln\left(
    \frac{\mathfrak{r}_s+\mathfrak{r}_{\mathrm{max}}}{\mathfrak{r}_s}
    \right)
    -\frac{\mathfrak{r}_{\mathrm{max}}}{\mathfrak{r}_s+\mathfrak{r}_{\mathrm{max}}}
    \right],
\end{eqnarray}
up to a maximum radius $\mathfrak{r}_{\mathrm{max}}$. On the other hand, the total electron mass encompassed in a spherically symmetric plasmic halo, characterized by the number density $N_2(r)$ in Eq.~\eqref{eq:N(r)-2}, can be obtained by doing an integration over the volume in the region $r_+<r<r_{++}$. This yields
\begin{eqnarray}\label{eq:MP}
    M_P &=& m_e \int_{r_+}^{r_{++}} N_2(r)\,4\pi r^2  \ed r\nonumber\\
    & =& \frac{4\pi m_e \omega_0^2}{105 r_+^2}\left[\right.
    3r_+^2\left(
    5r_+^7-35(b^2+\lambda^2)(r_{++}-r_+)
    -7r_+^5 r_{++}^2
 \right.\nonumber\\
    &&\left.\left. +2r_{++}^7\right)+7(6r_+^7-7r_+^5r_{++}^2+4r_+^2 r_{++}^5-3r_{++}^7)\right.\sigma^2\nonumber\\
    &&\left.-35(r_{++}^2-r_+^2)(r_+^3+2 r_{++}^3)\sigma^4 
    + 105(r_{++}-r_+)r_{++}^2\sigma^6\right].
    \end{eqnarray}
Solving the equation $M_P = N_{\mathrm{NFW}}$ for either of $\sigma$ or $\lambda$, one can get an estimation criteria, in which the plasmic surrounding can behave as a cold dark matter halo in the context of NFW description. Solutions to this equation however, although achievable, are rather complicated and do not have algebraic values. We instead, demonstrate the above criteria in a plot as in Fig.~\ref{fig:MOvsMNFW}.
The figure indicates more possible similarity between the electron plasma and the NFW cold dark matter, for the lower limits of $\sigma$ and $\lambda$. \\

In this section, We criticized the material distribution inferred from the two ansatzes for $n(r)$ and checked the criteria in which the PSK can be regarded as a dark matter halo. Now that the black hole's structure has been dealt with, in the next section, we try to illustrate mathematically the diameter of the black hole as it appears to an observer inside the causal region. This requires discussing the black hole's shadow.

%The deflecting light rays and those confined to the photon sphere, have in fact a direct relevance in the determination of the black hole's shadow. This is what we will take care of in the next section.

%%%%%%%%%%%%%%%%%%%%%%%%%%%%%%%%%%%%%%%%%%%%%%%%%%%
\section{Shadow of the black hole}\label{sec:shadowRadius}
The deflecting trajectories, governed by the angular equation of motion in Eq.~\eqref{eqmot}, can be divided into orbits of the first and second kind (respectively abbreviated as OFK and OSK). The former provides the well-known escape to infinity in terms of a definite deflection angle, whereas the latter results in falling onto the singularity \cite{Chandrasekhar:579245}. The OSK therefore result in the darkness of the sky for an observer who is observing the black hole. Hence, this observer encounters a dark disk which is the black hole's shadow. This shadow is surrounded by the photon trajectories following OFK. For this reason, it can be noticed by the observer. In this regard, the photon sphere is in fact the boundary of the shadow because it is the final possible limit, at which the photons can lie. The photon sphere is therefore unstable with respect to perturbations. This is essential in the determination of the shadow.

\begin{figure}[t]
    \centering
    \includegraphics[width=9cm]{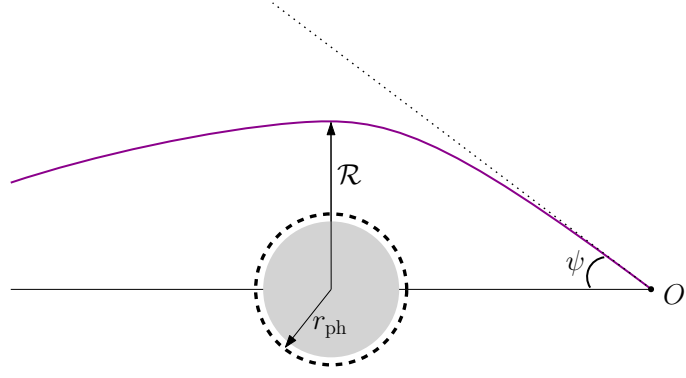}
    \caption{For an observer located at $O$, the angular diameter ($\psi$) of the black hole depends on the closest approach to the black hole. When $\mathcal{R}\rightarrow r_{\mathrm{ph}}$, then $\psi$ indicates the angular diameter of the shadow (here $\psi_{\mathrm{sh}}$).}
    \label{fig:shadowScheme}
\end{figure}

\begin{figure}[t]
    \centering
    \includegraphics[width=7.5cm]{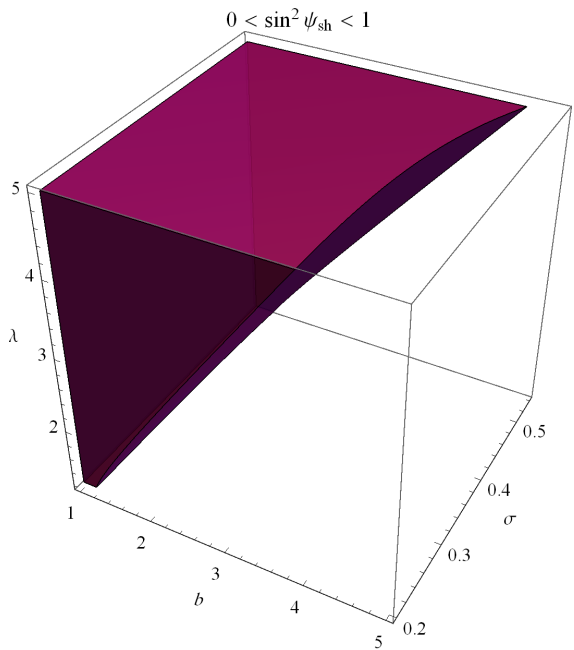}
    \caption{The allowed values of $b$ which satisfy the condition $0<\sin^2\psi_{\mathrm{sh}}<1$. The region has been plotted for $Q = 0.6$ and $r_O = 0.8$.}
    \label{fig:allowed-b}
\end{figure}

\begin{figure}[h]
    \centering
    \includegraphics[width=7cm]{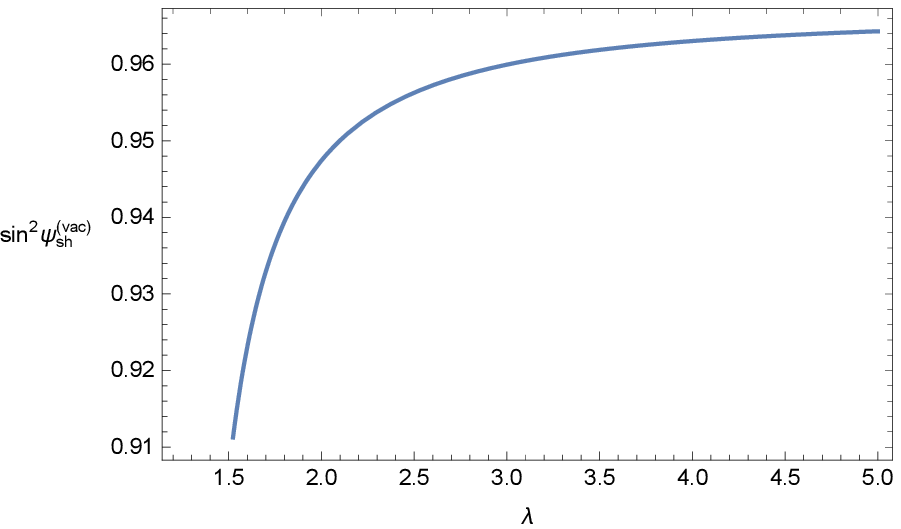}~(a)
    \includegraphics[width=7cm]{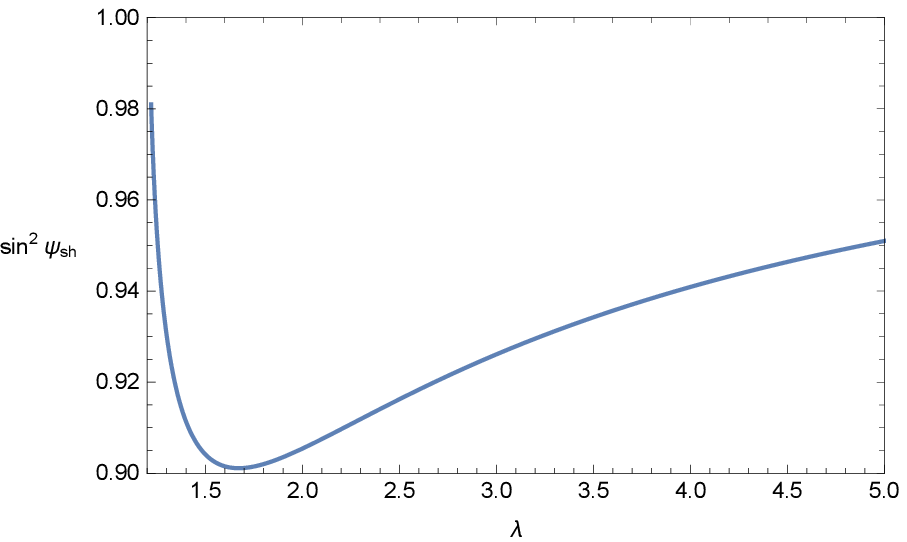}~(b)
    \includegraphics[width=9cm]{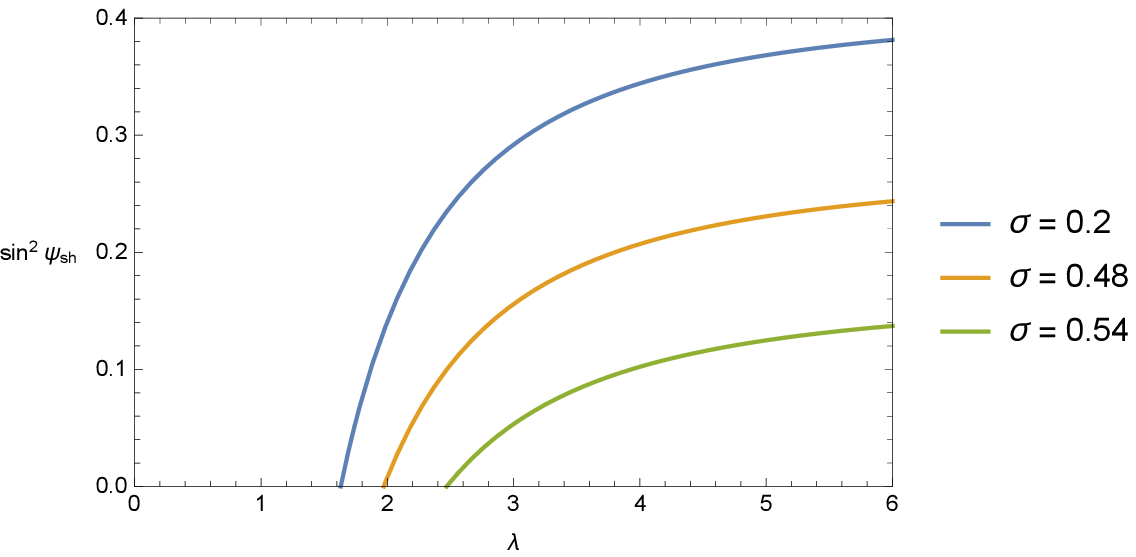}~(c)
    \caption{The radial diameter given in terms of $\sin^2\psi_{\mathrm{sh}}$, for (a) vacuum, (b) PFK and (c) PSK. The plots have been done for $Q = 0.6$, $r_O = 0.78$ and the impact parameter for the plots (b) and (c) has been taken as $b = 0.8$ (arbitrary length units have been considered).}
    \label{fig:sin2psi}
\end{figure}

To proceed, we calculate the angular diameter of the shadow, by considering an observer located outside the outermost photon sphere. Pursuing the method given in Ref.~\cite{Perlick:2015vta}, let us consider the scheme in Fig.~\ref{fig:shadowScheme}. The observer, located at the distance $r_O$, sends a light ray into the past at an angle $\psi$, which according to the line element in Eq.~\eqref{eq:generalmetric}, is given by
\begin{equation}\label{eq:psi-1}
    \psi = \arccot\left(
    \sqrt{\frac{1}{r^2B(r)}}\,\frac{\ed r}{\ed\phi}
    \right)|_{r=r_O},
    \end{equation}
which by means of Eqs.~\eqref{eqmot} and \eqref{ffunct}, becomes
\begin{equation}\label{psi-2}
    \psi = \arccot\left(
    \sqrt{\frac{h^2(r)}{h^2(\mathcal{R})}-1}
    \right)|_{r=r_O}.
\end{equation}
This can be recast as
\begin{equation}\label{psi-3}
    \sin^2\psi = \frac{h^2(\mathcal{R})}{h^2(r_O)}.
\end{equation}
Once the light rays have reached their final possible stable orbits at $r_{\mathrm{ph}}$, they indicate the outermost boundary of the black hole. Hence, the shadow can be determined by letting $\mathcal{R}\rightarrow r_{\mathrm{ph}}$ (see Fig.~\ref{fig:shadowScheme}). Accordingly, the corresponding angular diameter of the shadow is obtained as
\begin{equation}\label{eq:psiSh}
    \sin^2\psi_{\mathrm{sh}} = \frac{h^2(r_\mathrm{ph})}{h^2(r_O)}.
\end{equation}
Applying Eq.~\eqref{hfunct} we can calculate the above angle for the shadow. In the absence of plasma (i.e. for $h^2(r) = r^2/B(r)$), applying the radius in Eq.~\eqref{eq:rphoton-vacuum}, this angle becomes
\begin{equation}\label{eq:psiSh-vac}
    \sin^2\psi_{\mathrm{sh}}^{(\mathrm{vac})} = \frac{4 r_+^2 r_{++}^2 \left(r_O^2-r_+^2\right) \left(r_{++}^2-r_O^2\right)}{r_O^4 \left(r_{++}^2-r_{+}^2\right)^2}.
\end{equation}
For the PFK and PSK, discussed and analyzed in the previous sections, we get the following results:
\begin{itemize}
    \item From Eq.~\eqref{eq:ansatz1} we get
    \begin{equation}
            \sin^2\psi_{\mathrm{sh}} = \frac{r_+^2+r_{++}}{r_O^2+r_{++}},
    \end{equation}
for $r_{\mathrm{ph}} = r_+$ ($b = \sqrt{r_+^2+r_{++}^2}$).

\item From Eq.~\eqref{eq:ansatz2}, the angle in Eq.~\eqref{psi-3} becomes
\begin{equation}\label{eq:psi-general=2ndAnsatz}
    \sin^2\psi = \frac{b^2 r_+^2+\left(\sigma ^2+\mathcal{R}^2\right)^2 \left(r_{++}^2 \sigma ^2-r_+^2 \left(r_{++}^2-\mathcal{R}^2\right)\right)}{b^2 r_+^2+\left(r_O^2+\sigma ^2\right)^2 \left(r_{++}^2 \sigma ^2-r_+^2 \left(r_{++}^2-r_O^2\right)\right)}.
\end{equation}
Applying the condition in Eq.~\eqref{eq:psiSh} and the radius in Eq.~\eqref{eq:rph-secondn(r)}, the angular diameter of the shadow is obtain as
\begin{equation}\label{eq:psiShadow-mine}
    \sin^2\psi_{\mathrm{sh}} = \frac{\lambda^2 r_+^2}{\left(r_O^2+\sigma ^2\right)^2 \left(r_{++}^2 \left(r_+^2-\sigma ^2\right)-r_O^2 r_+^2\right)-b^2 r_+^2}.
\end{equation}

\end{itemize}
{Note that, not all values of $b$ are permitted to be possessed by the photons. This means that only certain photons with allowed impact parameters can identify the shadow. Such photons are those which could escape the black hole by passing the nearest possible distance (the critical distance) from it. According to the above relation, the condition $0<\sin^2\psi_{\mathrm{sh}}<1$ implies
\begin{equation}\label{eq:condition-on-b}
    b^2 < b_\mathrm{max}^2 -\lambda^2,
\end{equation}
in which
\begin{equation}\label{eq:bmax}
    b_\mathrm{max}^2  = \frac{\left(r_O^2+\sigma ^2\right)^2 \left(r_{++}^2 \left(r_+^2-\sigma ^2\right)-r_O^2 r_+^2\right)}{r_+^2}.
\end{equation}
This means that for every triplet $(\lambda,Q,\sigma)$, only photons satisfying the condition in Eq.~\eqref{eq:condition-on-b} can identify the shadow. In Fig.~\ref{fig:allowed-b}, a region has been plotted in which, the values of $b$ satisfy the above condition. Accordingly, and in Fig.~\ref{fig:sin2psi}, the angular diameters of the shadow have been plotted respectively for the vacuum, the PFK and the PSK.}
For all cases, no extremal black holes are observable. However, shadow of the black hole surrounded by the PFK, achieves its maximum angular diameter for the lower values of $\lambda$. This is while for the one corresponding to the PSK, $\psi_\mathrm{sh}$ tends to zero for same range of $\lambda$. This means that, this model of plasmic surrounding prohibits the shadow to appear to the observer, when the cosmological term in Eq.~\eqref{lapse} is dominant.\\

The discussion in this section, dealt with the way though which a charged Weyl black hole manifests itself to an observer residing in $r_+<r<r_{++}$. To demonstrate the shadow, it is usual to define some celestial coordinates which are obtained by doing a frame transformation from the curved background spacetime to the frame of a local observer (see for example the method of obtaining the shadow for rotating black holes in Refs.~\cite{Chandrasekhar:579245,Tsukamoto:2018} in vacuum and Ref.~\cite{Perlick:2017} in the presence of plasma. The latter is also applicable to the spacetimes which are not asymptotically flat. The case of static vacuum spacetime has also been investigated for example in Ref.~\cite{Singh:2018}). The shadow of the black hole under study, is completely symmetric and does not give more information other than those we have obtained so far. We therefore leave the discussion here and in the next section we bring the final notes and summarize the results.

%%%%%%%%%%%%%%%%%%%%%%%%%%%%%%%%%%%%%%%%%%%%%

%%%%%%%%%%%%%%%%%%%%%%%%%%%%%%%%%%%%%%%%%%%%%
\section{Conclusion}\label{sec:conclusions}

{The application of elliptic integrals in the determination of the deflection angle for the light rays propagating in a plasmic medium, surrounding a charged Weyl black hole, was the main aim of this paper. We calculated the equations of motion in connection with the plasmic energy density and refraction. Then by proposing two different ansatzes for the refractive index, we obtained analytical expressions for the light deflection which is the significance of the gravitational lensing caused by the black hole in the media. The solutions were given in terms of the elliptic integrals and for both kinds of plasma, we discovered that, depending on their energy and angular momentum, not all rays can contribute in the lensing process.}

Further, we demonstrated the particular way, through which, the radius of the photon sphere can be obtained. The photon sphere constitutes the closure of the final possible stable orbits around the black hole. In the first kind plasma, photons, regardless of their impact parameter, can form only one single photon sphere which depends only on black hole's characteristics. In contrast, the formation of photon sphere in the plasma of the second kind, depends directly on the test particles' energy and angular momentum and evolves in terms of the plasmic refraction. We continued our discussion by comparing the mass relation derived from the second kind plasma with that obtained from the NFW dark matter halo and demonstrated the extent of black hole properties, to which, these two could be similar in value. As the last concept, we considered the black hole's shadow and obtained its angular diameter in both cases of plasmic surrounding. The second kind plasma showed that not all photons can contribute in the formation of the shadow. We demonstrated this by plotting the angular diameter. 

{In conclusion, we highlight the importance of advanced mathematical methods in the investigation of light propagation in refractive media, when one is interested in inspecting the appearance of black holes to distant observers. For the cases studied here, we found that the impacts of plasma can make strong changes in the way the black hole is seen. In the present study, this became apparent in the demonstrated evolution of the deflection angles and the photon spheres.}

%%%%%%%%%%%%%%%%%%%%%%%%%%%%%%%%%%%%%%%%%%%%%%%%%%%%%

%%%%%%%%%%%%%%%%%%%%%%%%%%%%%%%%%%%%%%%%%%%%%%%%%%%%%%%%
\section*{Acknowledgements}
M. Fathi has been supported by the Agencia Nacional de Investigaci\'{o}n y Desarrollo (ANID) through DOCTORADO Grant No. 2019-21190382, and No. 2021-242210002. J.R. Villanueva was partially supported by the Centro de Astrof\'isica de Valpara\'iso (CAV).

\bibliographystyle{ieeetr}
\bibliography{Biblio_1.bib}
%\end{thebibliography}

\end{document}